\documentclass[10pt,showpacs,letterpaper,aps,twocolumn,nofootinbib,prx]{revtex4-1} 
\usepackage[draft]{hyperref}

\usepackage{amsthm,amsmath,amssymb,graphicx,comment} 
\usepackage{color}
\usepackage{bbold}
\usepackage{hyperref}
\usepackage[none]{hyphenat}

\newcommand{\ket}[1]{\left| #1 \right\rangle}
\newcommand{\bra}[1]{\left\langle #1 \right|}

\newcommand{\beq}{\begin{equation}}
\newcommand{\eeq}{\end{equation}}
\newcommand{\bea}{\begin{align}}
\newcommand{\eea}{\end{align}}

\begin{document}


\title{Verifying cross-Kerr induced number squeezing: a case study}
\author{David Schmid$^*$, Kevin Marshall, and Daniel F.V. James}
\affiliation{Department of Physics, University of Toronto, 60 St. George St.,
Toronto, Ontario, M5S 1A7, Canada}
\email{dfvj@physics.utoronto.ca}

\begin{abstract}
We analyze an experimental method for creating interesting nonclassical states by processing the entanglement generated when two large coherent states interact in a cross-Kerr medium. We specifically investigate the effects of loss and noise in every mode of the experiment, as well as the effect of `binning' the post-selection outcomes. Even with these imperfections, we find an optimal set of currently-achievable parameters which would allow a proof-of-principle demonstration of number squeezing in states with large mean photon number. We discuss other useful states which can be generated with the same experimental tools, including a class of states which contain coherent superpositions of differing photon numbers, e.g. good approximations to the state $\frac{1}{\sqrt{2}} (\ket{0}+\ket{20})$. Finally, we suggest one possible application of this state in the field of optomechanics.
\end{abstract}

\maketitle

\section{Introduction}

Thanks to advancements in the field of nonlinear optics, experiments with photon-photon interactions are rapidly becoming practical, leading to a flurry of related research. Achievable interaction strengths have increased dramatically in Kerr media \cite{exp, exp2, exp3, amir}, for which the index of refraction depends on an applied electric field intensity. When two light beams propagate through a Kerr medium, the electric field of each induces a phase shift on the field of the other, an effect known as the cross-Kerr effect. These interactions are being studied in various systems such as cavity QED, Rydberg atoms, ion traps, and superconducting qubits. A key goal in these experiments is to increase the interaction strength to the strong-coupling regime, in which various useful applications and experiments become feasible. For example, strong nonlinear effects enable universal gates for optical quantum computing \cite{QC, kerrgate} and the generation of squeezed states \cite{squeez, highly}, entangled states \cite{kerrgen, kerrgen2, highly, tian}, macroscopic superposition states\cite{tian, highly, macro, weakkerrgen2}, and resource states for quantum information protocols \cite{QIP1, weakkerrloss1, weakkerrgen2, highly}. In the microwave regime, superconducting qubit technologies already allow for some of these basic protocols to be implemented \cite{exp3}. 

Even with the weak Kerr interaction currently achievable, many similar protocols have been suggested or demonstrated in the optical regime. These schemes typically involve interactions with at least one large coherent state, which compensates for small per-photon interaction strengths. Using post-selection, various types of measurement, and feed-forward, schemes using weak nonlinearities have also been proposed for enabling quantum computation \cite{QC2, kerrgate}, creating specific quantum gates \cite{kerrgate}, and generating macroscopic superposition states, entangled states, and other nonclassical states \cite{tian, weakkerrloss1, weakkerrgen2}.

Many of the non-Gaussian states whose creation have been suggested (and which we suggest below) have applications as resource states in various quantum information settings such as teleportation \cite{tele}, metrology \cite{metrol}, non-demolition photon number detection \cite{ND}, off-line resources for creation of quantum computing gates \cite{cat, resource2}, and entanglement generation, broadcasting, and distillation \cite{entang, distill, QIP1}. Some of the non-Gaussian states we propose here are well-studied, e.g. number-squeezed states \cite{mandel,teich89} (a state for which the variance in the photon number distribution is less than the mean) or Fock states, but in new parameter regimes (e.g. large mean photon number); others are altogether novel. Given that the interaction strengths required for some of these proposals are experimentally accessible, the primary remaining obstacles are loss and noise, which rapidly degrade the quality of the desired states or gates. As such, considerable effort has been spent characterizing the robustness of these protocols \cite{weakkerrloss1, weakkerrgen2}. 

This paper consists of two parts: a specific experimental proposal and an exploration of idealized capabilities within the same setup. Like many of the above cited works, we rely on the entangled state generated by two modes interacting in a weak cross-Kerr medium. We propose a post-processing which, in the ideal case, allows the harnessing of this entangled state for the creation of most of the resource states above. In less ideal circumstances, such as those currently achievable in the laboratory, we use our scheme to generate number-squeezing for states of large mean photon number. This proof-of-principle experiment would demonstrate the capabilities of weak cross-Kerr experiments as well as paving the way forward towards the aforementioned proposals.

Several features of our proposal and our model improve upon past research. First, our experimental proposal directly uses the capabilities and limitations inherent in current experiments, specifically those in \cite{amir}, and considers a natural follow-up experiment given the capabilities demonstrated therein. Second, our model treats both loss and noise realistically; the latter is typically not considered, yet is found to be extremely detrimental to the verification of nonclassical effects and the generation of many nonclassical states. Third, we demonstrate that the number-squeezing is robust to the binning of outcomes required in such experiments (with post-selection on continuous variables). These three advantages make our model directly applicable to experiments that are currently being attempted in cross-Kerr laboratories. Specifically, our results elucidate the various trade-offs between experimental design parameters, and as such can be used to guide in the design of an apparatus. These trade-offs are summarized at the end of section 5. Finally, we propose a novel class of states and suggest its use to augment interaction strengths in optomechanical experiments.

\section{Theoretical scheme}
\label{sec:basictheory}

\begin{figure}[htbp]
    \centering
    \includegraphics[width=\columnwidth]{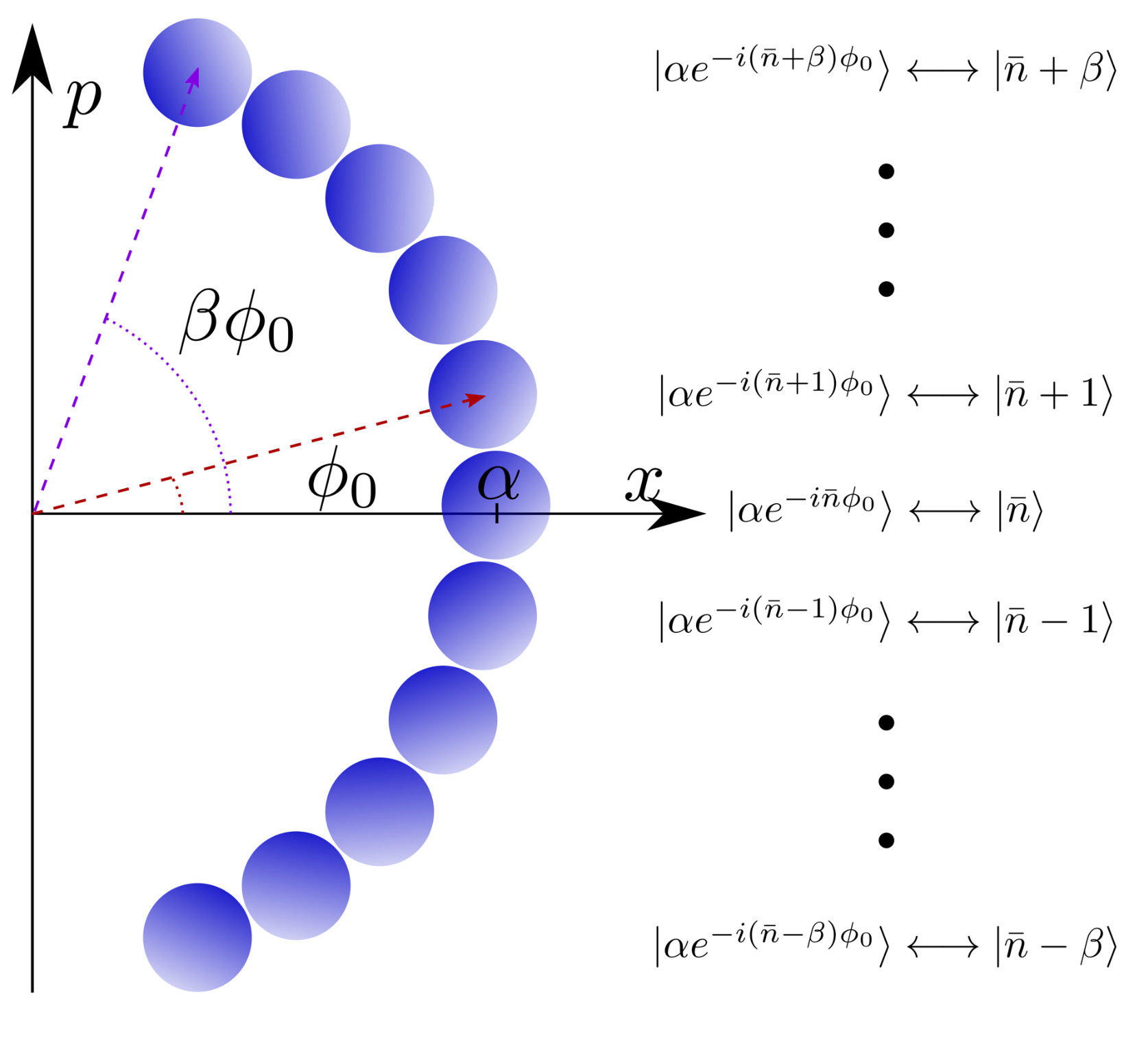}
    \caption{(color online) Rough visualization of number-phase entangled state: each number state of the signal (noted in the kets on the right) causes a different phase shift, leading to the rotations of the probe state (noted in the kets on the left), pictured here in the phase space of the probe.  }
    \label{entang}
\end{figure}

Our proposed experimental setup uses existing technological components, many of which recently enabled the first observation of phase shifts caused by single photon intensities \cite{amir}. Experimental details of the sources, Rubidium-based cross-Kerr medium, and measurements can be found in Section \ref{sec:basicexp} or in \cite{amir}. For our purposes, two coherent states  $\ket{\psi_0} = \ket{\beta} \ket{\alpha} $ with real amplitudes $\alpha$ and $\beta$ interact in a nonlinear medium via a Hamiltonian $\chi \hat{n}_a \hat{n}_b $, inducing an evolution operator $e^{-i \phi_0 \hat{n}_a \hat{n}_b}$ and resulting in the final state 
\begin{align}\label{eq:ideal}
e^{-i \phi_0 \hat{n}_a \hat{n}_b} \ket{\psi_0} &= e^{-i \phi_0 \hat{n}_a \hat{n}_b}e^{-\frac{1}{2}|\beta|^2} \sum_{n = 0}^{\infty} \frac{\beta^n}{\sqrt{n!}} \ket{n} \ket{\alpha}  \nonumber\\
&= e^{-\frac{1}{2}|\beta|^2} \sum_{n = 0}^{\infty} \frac{\beta^n}{\sqrt{n!}}  \ket{n} \ket{\alpha e^{-i \phi_0 n}},
\end{align} 
where $\phi_0$ is the phase shift caused by a single photon.  Clearly, the photon number in the first mode (the signal $\beta$) is entangled with the phase of the second mode (the probe $\alpha$); this number-phase entanglement is discussed in depth in \cite{tian}, and used to create appreciable micro-macro states even when the interaction strength is small. The entangled state can be visualized as in Figure \ref{entang}.

If one could perform a measurement which revealed the phase of the coherent state in the second mode, then one could infer approximately the number of photons in the first mode. Post-selecting on phase outcomes within a narrow range about some mean value, one is left with a number-squeezed state in the first mode, with photon numbers distributed in a narrow range about a mean value (corresponding to the chosen mean phase rotation). Intuitively, we can understand this squeezing via Bayesian inference: consider an initial number distribution of a coherent state with an average photon number of $\bar m$        \begin{equation}P(m) = \frac{e^{-\bar m} \bar{m}^m}{m!} \approx \frac{1}{\sqrt{2\pi\bar m}}e^{\frac{-(m-\bar{m})^2}{2\bar{m}}}\end{equation}       and a phase measurement likelihood function        \begin{equation}P(\phi | m) = e^{\frac{-(\phi-m \phi_0)^2}{2\sigma^2}} = e^{\frac{-(m - \phi/\phi_0)^2}{2(\sigma/\phi_0)^2}},\end{equation} where $\sigma$ is the precision of the phase measurement.      The photon number distribution given a phase measurement with outcome $\phi$, then, becomes        \begin{equation}P(m|\phi) = e^{\frac{-(m - \phi/\phi_0)^2}{2(\sigma/\phi_0)^2}}  e^{\frac{-(m-\bar{m})^2}{2\bar{m}}},\end{equation}       a Gaussian distribution with a decreased width of \begin{equation}\frac{\bar{m} \sigma^2/\phi_0^2}{\bar{m} + \sigma^2 / \phi_0^2}.\end{equation} From this reasoning, we see that the strength of the squeezing scales with the ratio between the phase measurement  precision and the phase shift per photon, as one might expect intuitively. In the ideal case, the above logic would allow for creation of a number-squeezed state given any realistic parameters, such as those in the experiment by Feizpour et al \cite{amir}. However, a realistic experiment must contend with noise, imperfect measurements, extreme sensitivity of entanglement to loss, and a trade-off between post-selection success rates and degree of squeezing. We now determine the feasibility of such an experiment with all relevant imperfections and with realistic parameters. 

\section{Experimental scheme}
\label{sec:basicexp}

The experimental apparatus on which our proposal is based can be found in \cite{amir}; the only missing feature is a standard measurement of second-order coherence, $g^{(2)}$, which requires no new optical components. In \cite{amir}, a Kerr medium of rubidium atoms is used to mediate an effective interaction between two coherent states. The presence of a signal beam modifies the index of refraction of a cold cloud of rubidium atoms trapped in a magneto-optical trap; a second beam then acquires a modified phase shift proportional to the signal intensity. The proportionality constant $\phi_0$ gives the phase shift caused by a single signal photon; in this experimental arrangement, values of $\phi_0$ near 20 $\mu$rad have been demonstrated. This effect is strongest on resonance, where absorption dominates and electromagnetically-induced transparency (EIT) is necessary to minimize dissipation. However, EIT guarantees a linear response (i.e. a phase shift proportional to signal intensity) only for relatively weak signal fields. For the experiment considered here, we require $\beta \lesssim 70$; for larger values, the phase shift fails to be linear (and eventually begins to decrease). There is also a limitation on the probe intensity to roughly $\alpha \lesssim 70$; above this intensity, residual probe absorption due to imperfect EIT rapidly depletes the number of trapped rubidium atoms. After the interaction, the amplitude and phase change in the probe were measured with a phase error of roughly 20 milliradians. Finally, a $g^{(2)}$ measurement could easily be implemented in the signal arm. The error in this $g^{(2)}$ measurement is dominantly from dark counts, which are largely proportional to the crosstalk between signal and probe beams as well as scattering from the rubidium atoms. In such a setup, net losses of under $50\%$ (due primarily to inefficient collection optics and detectors) can easily be achieved in each arm. We elaborate on these rough parameter choices below, and more details on these effects and the experimental setup and capabilities can be found in \cite{amir}.

\section{Analysis of number squeezing}

\begin{figure}[htbp]
    \centering
    \includegraphics[width=\columnwidth]{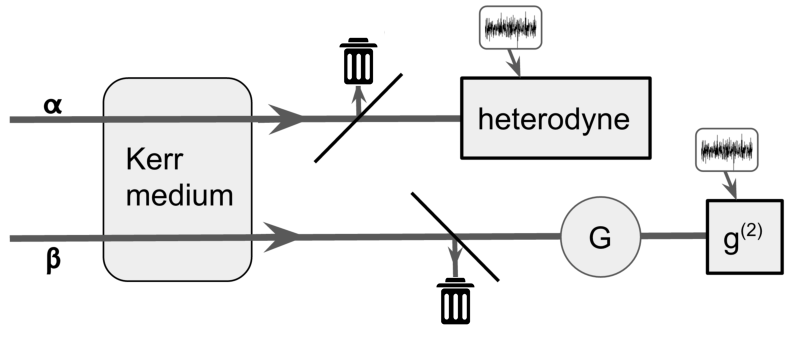}
    \caption{Model of our experimental proposal. Two modes $\alpha$ and $\beta$ are entangled via a cross-Kerr medium. Loss in each mode is modeled by a beamsplitter coupling to the environment, which is then traced over (denoted by a trash can). The probe mode undergoes a noisy heterodyne measurement, after which we can approximate our signal state as a Gaussian (schematically represented by `G'). Finally, the $g^{(2)}(0)$ of our state is measured in the presence of dark counts. These dark counts and the heterodyne noise are illustrated as random inputs to our measurements.}
    \label{schem}
\end{figure}

The ideal entangled state after our Kerr medium is given by \eqref{eq:ideal}.
Our goal is to implement a post-selection on the second mode, followed by a measurement of the squeezing of the first mode, in the presence of all relevant and realistic experimental imperfections. Figure \ref{schem} shows a schematic of our proposed experiment. The ideal state, from \ref{eq:ideal}, undergoes loss in both arms, followed by a noisy heterodyne measurement and a noisy $g^{(2)}(0)$ measurement on the probe and signal, respectively. The precise implementation of each of these steps is discussed below. 


On the second mode, we expect loss due to imperfect optics and detectors. We model this loss by coupling the mode of interest via a beamsplitter to an auxiliary environment mode. Because this single auxiliary mode is never observed in practice, we immediately trace over it, leaving both remaining modes in a partially mixed state. It is this two-mode state on which we do a noisy phase measurement as follows.

Experimentally, the phase rotation of a state with respect to a reference can be measured by heterodyne or homodyne detection.  An idealized heterodyne measurement is well described by a positive-operator valued measure $\{\frac{1}{\pi}\ket{\delta}\bra{\delta}\}$ consisting of sub-normalized projectors onto the set of coherent states $\{\ket{\delta}\}$ \cite{heter}. The updated state after outcome $\delta'$ and tracing out the second mode is  $\bra{\delta'} \bar{\rho} \ket{\delta'}$. Modeling our heterodyne measurement, then, requires projecting the second mode of the joint state onto a set of coherent states. Each outcome corresponds to a rotation angle of the state in the second mode in phase space, corresponding to a specific number state in the first mode, as in Figure \ref{entang}. Next, we implement a smooth post-selection envelope by probabilistically sampling the outcomes; a given outcome $\delta'$ is included in our fixed post-selection value $\delta$ with an unnormalized probability $\exp[-|\delta'-\delta|^2/2 \epsilon^2]$. We are only interested in one `bin': the specific $\delta$ which corresponds to phase shifts of mean value $\phi_0$. The optimal post-selection envelope width $\epsilon$ depends on the goal: a large width implies a higher probability of post-selection success and hence more data, but also implies a washing out of the number-squeezing. \footnote{However, if high data rates are crucial, a more sophisticated experiment could use feedback to keep more of the signal photons: one would simply apply a displacement operator whose magnitude was a function of the phase measurement's outcome, such that the mean photon number in all shots was identical after the displacement. This would allow one to include more of the measurement results while suffering only a small decrease in squeezing \cite{highly}.}

The primary imperfection for heterodyne measurement is random errors on each phase outcome. Errors arising from fluctuations (e.g. in laser power or detector shot noise) are well approximated by Gaussian noise due to the central limit theorem, and further Gaussian randomness arises from the inherent angle uncertainty of a coherent state. Any such Gaussian random uncertainty can be modeled simply by averaging over a Gaussian set of projectors, which leads to precisely the same form of washing-out as the above Gaussian post-selection. As such, both effects can be accounted for by averaging the original post-selection $\bra{\delta'} \bar{\rho} \ket{\delta'}$ over a single Gaussian envelope
\begin{equation} 
\frac{1}{2 \pi \Delta^2} \int^{\infty}_{-\infty} e^{\frac{-|\delta'-\delta|^2}{2 \Delta^2}} \bra{\delta'} \bar{\rho} \ket{\delta'} d^2\delta',
\end{equation}
whose width is given by the quadrature sum of the widths from sampling and from phase noise. For some standard implementations in the lab, it has been observed \cite{matin} that the technical phase uncertainty $\delta \phi$ in a heterodyne measurement is approximately fixed. This implies that the Gaussian width accounting for the phase noise must scale linearly with the input state amplitude, say $\gamma$, so that the total Gaussian envelope width $\Delta$ is given by $\sqrt{\epsilon^2 + (\gamma \tan{(\delta \phi /2)})^2}$.

After these practical imperfections are accounted for, the state in the first mode is significantly less squeezed than in the ideal situation. In fact, one can show that, for parameters similar to those in \cite{amir}, the resultant state is approximately Gaussian: it deviates only slightly from a quadrature-squeezed state. In a future, improved experiment, this approximation breaks down strongly, and the state of the second mode in phase space stretches roughly into a crescent shape. Therefore, further calculations are simplified greatly by assuming the state in the first mode is indeed Gaussian, and can thus be fully characterized by its covariance matrix \textbf{M} and displacement vector \textbf{d}.  We can equivalently characterize our state by expressing it in Williamson form \cite{Williamson36} and extracting the squeezing parameter $r$ and the thermal variance $V$.  This form is a convenient representation for quadrature-squeezed states; the variance $V$ quantifies how close a state is to saturating the minimum possible uncertainty, while the squeezing factor $r$ quantifies the trade-off between the the two quadratures' uncertainties. The displacement vector can be similarly calculated, but is unimportant to us, as discussed below.

Manipulations on our covariance matrix are generally easy and computationally efficient. For example, we model loss as before with a beamsplitter coupling to an auxiliary vacuum mode, which in the covariance matrix transforms as $\textbf{M} \rightarrow \textbf{SMS}^T$ and the displacement vector transforms as $\textbf{d} \rightarrow \textbf{Sd}$, where the symplectic matrix $\textbf{S}$ corresponds to a beamsplitter between the two modes \cite{weedbrook}. 
However, subtleties arise when one includes detector dark counts in a calculation of the experimentally measured squeezing.

To characterize the number-squeezing induced on the first mode, we wish to calculate the second-order coherence function $g^{(2)}(\tau) = \text{Tr}[\rho \hat{a}^{\dag}(t)\hat{a}^{\dag}(t + \tau) \hat{a}(t + \tau) \hat{a}(t)] \Big/ \text{Tr}[\rho \hat{a}^{\dag}(t) \hat{a}(t)]^2$. For states with totally uncorrelated photon arrival times, this quantity is uniformly 1; we say the photons are antibunched if $g^{(2)}(\tau)>g^{(2)}(0)$ and bunched if $g^{(2)}(\tau)<g^{(2)}(0)$.  For all states, if $\tau$ is longer than the state's coherence length, then the $g^{(2)}(\tau)$ tends to 1.  When $g^{(2)}(0)<1$ the photon number statistics are sub-Poissonian and when $g^{(2)}(0)>1$ they are super-Poissonian  \cite{zou,gerry}. In this experiment, we need only consider $g^{(2)}(0)$, since it is most easy to verify any squeezing that might be present when $\tau = 0$. For a state with less than one photon on average (which will be all we require), this can be measured experimentally by splitting the input state on a beamsplitter and using photodetectors in each resultant mode to find $g^{(2)}(0) = P_{cc} \Big/ P_{ca}P_{ac}$, with singles $P(ca) \text{ and } P(ac)$ and coincidence $P(cc)$ counts. Here, $c$ means a click on a detector, and $a$ indicates any outcome, i.e., click or no click. Essentially, $a$ corresponds to a trace over the corresponding mode. The letters are ordered to indicate mode one first and mode two second. Calculating the probabilities for these various events from our covariance matrix, in the presence of noise, involves some conceptual subtlety, which we now elaborate upon.

Although in principle we could construct a density matrix for the state defined by $\mathbf{M}$ and $\mathbf{d}$, it is easier to directly express the state using the Wigner function \cite{gerry}: 
\begin{align}
W(\mathbf{r})&=\frac{e^{-(\mathbf{r}-\mathbf{d})^T \mathbf{M}^{-1} (\mathbf{r}-\mathbf{d})}}{\pi^2 \sqrt{det(\mathbf{M})}} ,
\end{align}       
where $\mathbf{r} = (x,p)^T$. To calculate the probability of collapse, up to a constant factor, onto a pure state in the Wigner representation, we can simply take the inner product of the functions by integrating their product over all phase space. We further simplify these calculations by looking only at outcomes which correspond to Gaussian states and thus Gaussian Wigner functions: namely, the outcome corresponding to finding the state in the vacuum (labeled `n' for no click), or the trivial outcome corresponding to tracing out a mode (labeled `a' for any outcome).  Dark counts are defined as that state of affairs which guarantees the detector will click and occur at a rate parametrized by the experimentally testable parameter $P_d$, given a fixed time window over which the detectors count. Thus each mode contains a dark count ($d$) with probability $P_d$ and does not contain a dark count ($\bar{d}$) with probability $P_{\bar{d}}$. All terms in the experimental version of $g^{(2)}(0)$ can be written entirely in terms of these easy-to-calculate conditional probabilities: $P(an|\bar{d}\bar{d})$, $P(na|\bar{d}\bar{d})$, and $P(nn|\bar{d}\bar{d})$. We have 
\begin{align}
P(ca) &= 1- P(na|\bar{d}\bar{d})P(\bar{d}) \nonumber\\
P(ac) &= 1- P(an|\bar{d}\bar{d})P(\bar{d})\nonumber\\
P(cc) &= P(ac) - P(na|\bar{d}\bar{d})P(\bar{d}) - P(nn|\bar{d}\bar{d})P(\bar{d})^2.
\end{align}
Therefore, we can extract all three necessary components of the experimental $g^{(2)}(0)$ formula via calculations of probabilities requiring only integrals of Gaussian products defined by the vacuum, the trace, and the final state of the first mode.


Our simulations show (and it is easy to calculate) that one can gain an appreciable decrease in $g^{(2)}(0)$ for some optimal displacement (typically to around 0.5 photons on average). It is for this reason that calculating the displacement vector for our signal state was unimportant; in every case we displace our signal state to its optimal value just before measuring its $g^{(2)}(0)$. Experimentally, this displacement requires combining the signal state with a strong coherent state on a highly reflective beamsplitter. If the beamsplitter reflectivity or laser power is not ideally calibrated, the $g^{(2)}(0)$ will be shifted slightly from the ideal value; however, because $g^{(2)}(0)$ has a minimum at our desired displacement, this change to $g^{(2)}(0)$ will be negligible.

\section{Simulated experimental results}

\begin{figure}[htbp]
    \centering
    \includegraphics[width=\columnwidth]{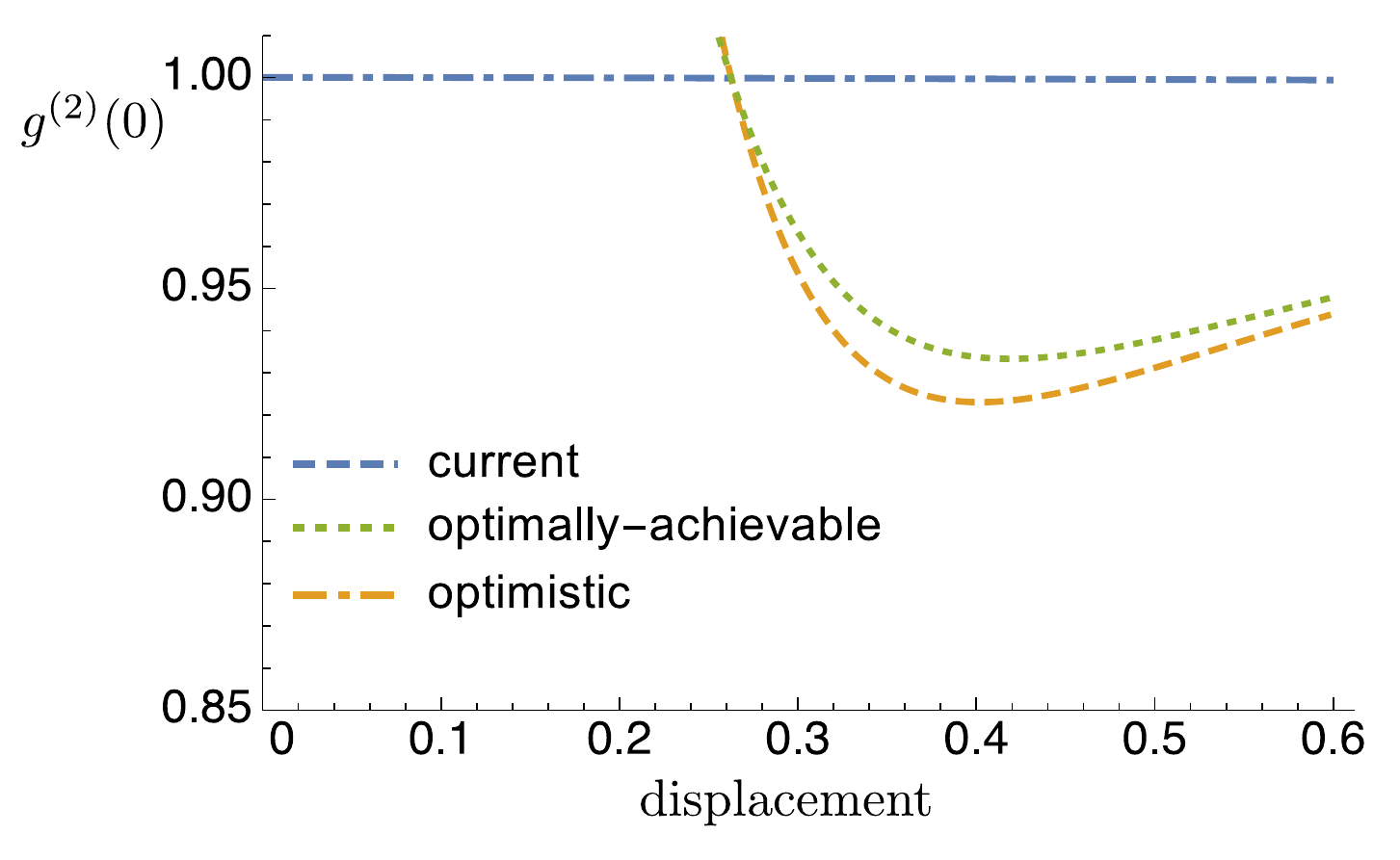}
    \caption{(color online) With currently achievable parameters (blue dot-dashed line), the $g^{(2)}(0)$ is virtually indistinguishable from that of a Poissonian state ($g^{(2)}(0) = 1$). With slightly more optimistic, yet still achievable parameters (green dotted line), the $g^{(2)}(0)$ dips to 0.93 near an optimal displacement of approximately 0.41. For optimistic parameters (orange dashed line), the $g^{(2)}(0)$ achieves a minimum value near 0.92 for an optimal displacement of approximately 0.395. Each set of parameters corresponds to a set introduced in Table~\ref{params}.}
    \label{currentoptim}
\end{figure}

With our model in hand, we analyze the predicted values that would be observed experimentally for our cross-Kerr induced number-squeezing by plotting the experimental $g^{(2)}(0)$ while varying other parameters. Potentially variable parameters include the loss in the first mode ($\eta$) and the second mode  ($\nu$), phase noise error  ($\delta \phi$), post-selection envelope size  ($\epsilon$), detector dark counts  ($P_d$), per-photon phase shift ($\phi_0$), and the size of our two incident coherent states, the probe ($\alpha$) and signal ($\beta$).  We first start our model with experimental parameters given by the \emph{Current} set in Table~\ref{params},  similar to those achievable in \cite{amir}, leading to a $g^{(2)}(0)$ versus displacement as shown in Figure \ref{currentoptim}.  Clearly, only a tiny amount of number-squeezing is created, such that any imperfections not considered here (e.g. uncompensated drifts) would likely wash out any verifying measurements. From here, we tweak the parameters within a reasonable range to determine their sensitivity in search of mildly optimistic parameters ranges which significantly improve on the squeezing shown in Figure \ref{currentoptim}. 

\begin{table}
	\centering
	\begin{tabular}{ l  c  c  c c c c c c }
   & $\eta$ & $\nu$ & $\delta\phi$ & $\epsilon$ & $P_d$ & $\phi_0$ & $\alpha$ & $\beta$\\ \hline\hline
  Current &  0.5 & 0.5 & 0.02 & 0.3 & 0.1 & 0.00002 & 50 & 50 \\
  Optimally-achievable & 0.5 & 0.5 & 0.01 & 0.3 & 0.001& 0.00002 & 70 & 70\\
  Optimistic  & 0.5 & 0.5 & 0.01 & 0.3 & 0.0001& 0.00002 & 70 & 70\\
  \hline\hline
\end{tabular}
\caption{The different sets of experimental parameters we consider.  The \emph{Current} set is achievable today, e.g. in the experiments of \cite{amir}. The  \emph{Optimistic} set requires improvements to current experimental designs. We therefore suggest the \emph{Optimally-achievable} set, which is nearly as easy to achieve as the \emph{Current} set, but generates nearly as much squeezing as the \emph{Optimistic} set.  }\label{params}
\end{table}

\begin{figure}[htbp]
    \centering
    \includegraphics[width=\columnwidth]{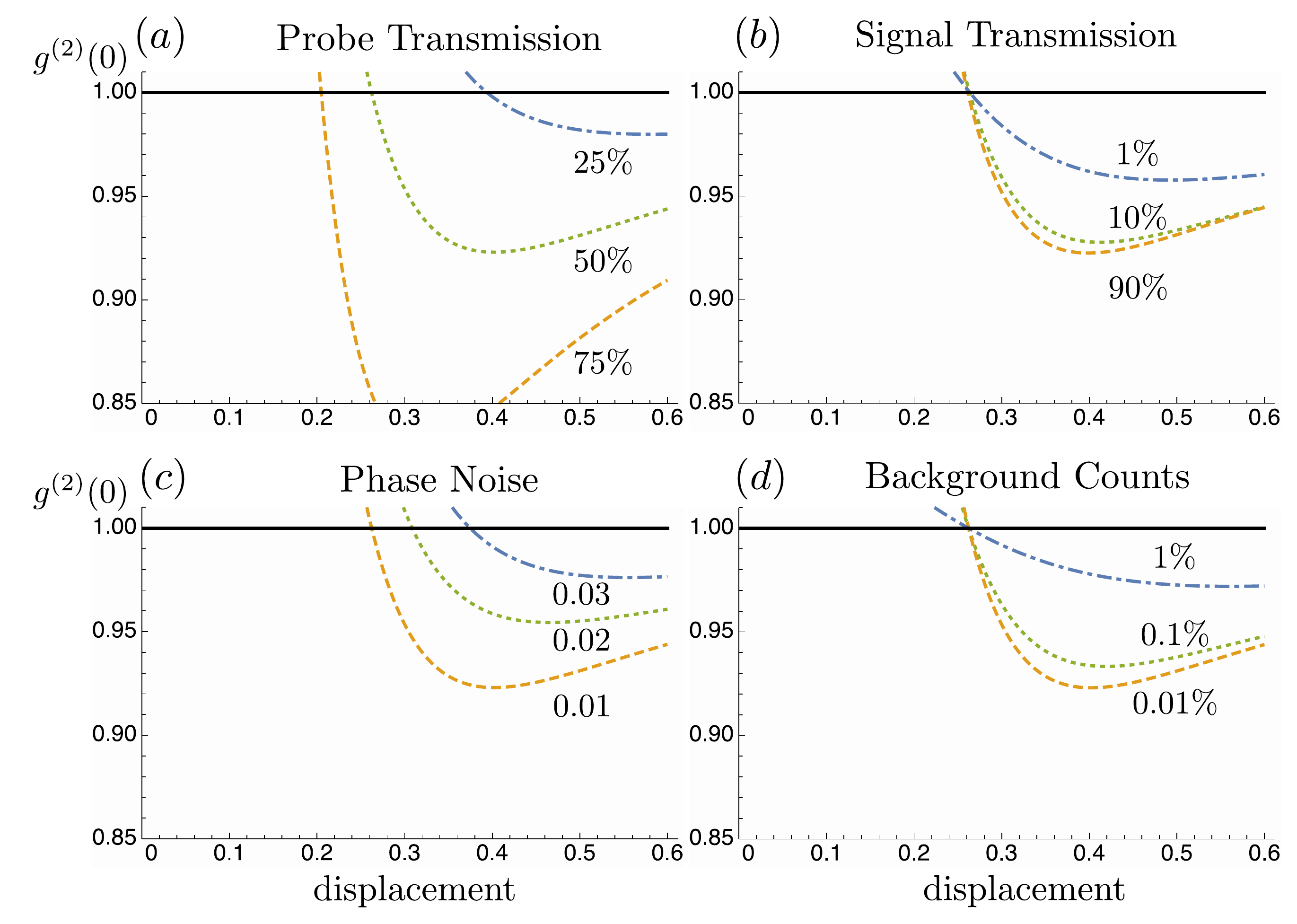}
    \caption{(color online) \textbf{(a)} Plot of $g^{(2)}(0)$ versus displacement for three values of probe transmission. \textbf{(b)} Plot of $g^{(2)}(0)$ versus displacement for three values of signal transmission. \textbf{(c)} Plot of $g^{(2)}(0)$ versus displacement for three values of the heterodyne phase noise. \textbf{(d)} Plot of $g^{(2)}(0)$ versus displacement for three background count rates. As argued above, probe transmission, phase noise, and background counts are all sensitive parameters, while signal transmission is quite insensitive.  All fixed parameters in the plots are given by the \emph{Optimistic} set of parameters defined in Table~\ref{params}.}
    \label{joined}
\end{figure}

As an optimistic baseline, we consider the parameters described by \emph{Optimistic} given in Table~\ref{params}. For these optimistic parameters, we see that the $g^{(2)}(0)$ dips below 0.93, a value which is easily distinguishable from 1. The post-selection success probability (discussed in depth at the end of this section) was in this case 15.2\%, and the optimal displacement corresponds to a mean photon number of approximately 0.395. For all realistic parameter variations about these values, the $g^{(2)}(0)$ dips below 1, but diverges for vanishing displacement, so there always exists an optimal displacement for which $g^{(2)}(0)$ reaches a minimum. As claimed, the sub-Poissonian behavior is strongly suppressed by entanglement-degrading loss in the probe arm, as shown in Figure \ref{joined}. In this same figure, part b, we see that low efficiency in the signal arm is less detrimental, as it occurs \textit{after} the post-selection eliminates all entanglement. \footnote{We assume most of the inefficiency occurs from detector inefficiencies.} In fact, the only requirement on the transmission in the signal arm is that it be at least an order of magnitude higher than the dark count probability. Noise on our phase measurement is also a sensitive parameter, as argued above and shown in Figure \ref{joined}. Similarly, background counts on the detectors used to measure the $g^{(2)}(0)$ of mode 1 are extremely detrimental, as verified in Figure \ref{joined}. In fact, this is the most vital parameter requiring experimental improvements. At least one or two orders of magnitude reduction in the count rate is necessary for any observation of squeezing; this is the (only) limiting factor in our proposed experiment. Other experimental apparatuses would likely not suffer from such large dark counts, as the current experimental rates are largely determined by a two-photon scattering process specific to the rubidium atoms in \cite{amir}.

In contrast, the $g^{(2)}(0)$ is insensitive to the interaction strength and the initial probe and signal state sizes. One might have expected these parameters to be very important, and indeed they are for an ideal experiment, for which an increase in any of the three leads to larger number-phase entanglement, which leads to a larger squeezing. However, in a loss-dominated experiment, it also leads to a larger leakage of information to the environment, which washes out almost all gains. For a setup less dominated by loss, these parameters become quite sensitive again. For our purposes, however, we can decrease their values (e.g. in order to decrease background counts from scattered probe light) with little cost; the only detriment in this case is a slight increase in sensitivity to background counts.\footnote{An additional limitation is imposed by the theoretical limit $\text{min}[\delta \phi] = \frac{1}{2|\alpha|}$, which becomes important once other sources of phase noise are removed and for smaller $\alpha$ than considered here.} 

Based on all of these observations and through discussion with those in Ref.~ \cite{matin}, we propose an optimally achievable set of parameters, denoted \emph{Optimally-achievable} in Table~\ref{params}, to create and verify number-squeezing on a large coherent state in a realistic cross-Kerr experiment. This set of parameters is at the limit of experimental accessibility.

\begin{figure}[htbp]
    \centering
    \includegraphics[width=\columnwidth]{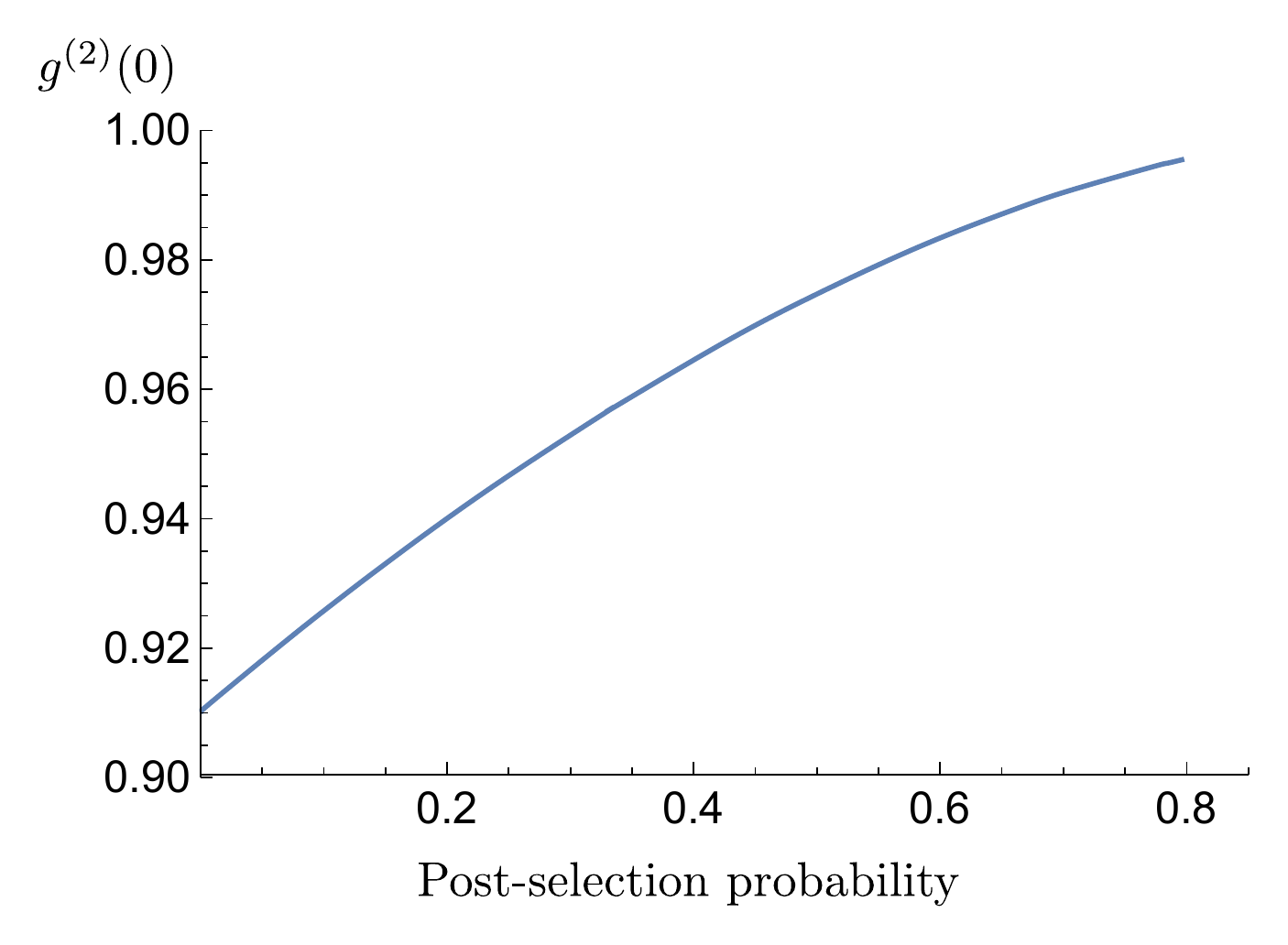}
    \caption{(color online) For increased probabilistic sampling width $\epsilon$, the probability that a given outcome passes the post-selection increases, while the number squeezing is slowly washed out. The parameters utilized in this plot are those defined in the optimally achievable set.}
    \label{g2prob}
\end{figure}

Lastly, we must verify that the $g^{(2)}(0)$ is still distinguishable from 1 after the degradation caused by the required binning of heterodyne outcomes. In Figure \ref{g2prob}, we plot the trade-off between $g^{(2)}(0)$ and the probability that a post-selection event succeeds, given the set of optimally achievable parameters. At small success rates, the trade-off is almost linear, and for post-selection success rates of roughly 10\% the $g^{(2)}(0)$ decreases from its ideal value by 16.7\%. This degradation should be manageable without using more complicated feedback techniques like those mentioned above. Obviously, an experimentalist would ultimately choose the smallest probabilistic sampling width $\epsilon$ which did not require prohibitively long data-taking sessions, depending on (for example) drift rates.

Finally, we roughly summarize the various trade-offs and (in)sensitive parameters uncovered by our model. These points (and the corresponding plots above) should be useful for designing related experiments.
\begin{itemize}
  \item Probe transmission, phase noise, and background counts are all extremely sensitive parameters, while signal transmission is not very sensitive.
  \item The strength of the Kerr nonlinearity and signal and probe magnitudes are not extremely sensitive parameters in our proposed experiment, counter to what intuition might suggest. The primary benefit to increasing their strength is to mitigate the detrimental effects of loss and noise.
  \item The required binning (or probabilistic sampling) of post-selection outcomes degrades the number squeezing, but to a manageable degree.
\end{itemize}
	
\section{Theoretical state-creation}

In this section we discuss more generally what types of interesting states can be created by processing the entangled state created in our interaction medium. Above, we mentioned micro-macro entangled states which can be harnessed for continuous-variable quantum information processing, and we discussed creation of bright number-squeezed states. In the simplest case, one can imagine increasing the per-photon phase shift until the number-squeezing discussed herein becomes ``complete'', in the sense that the final state is left in a Fock state with a large and precise photon number. These fundamentally interesting states are known to be useful for such tasks as optimal quantum communication \cite{comm}, error correction \cite{bosonic}, and interferometry \cite{interf}, and likely will find use in future experiments such as boson sampling. This idea requires too ambitious a set of parameters to be carried out with current technology. For smaller interaction strengths, however, sub-Poissonian states with large photon number could be of practical use, e.g. for absorption metrology \cite{metrol}. 

\begin{figure}[htbp!]
\centering
\includegraphics[width=\columnwidth]{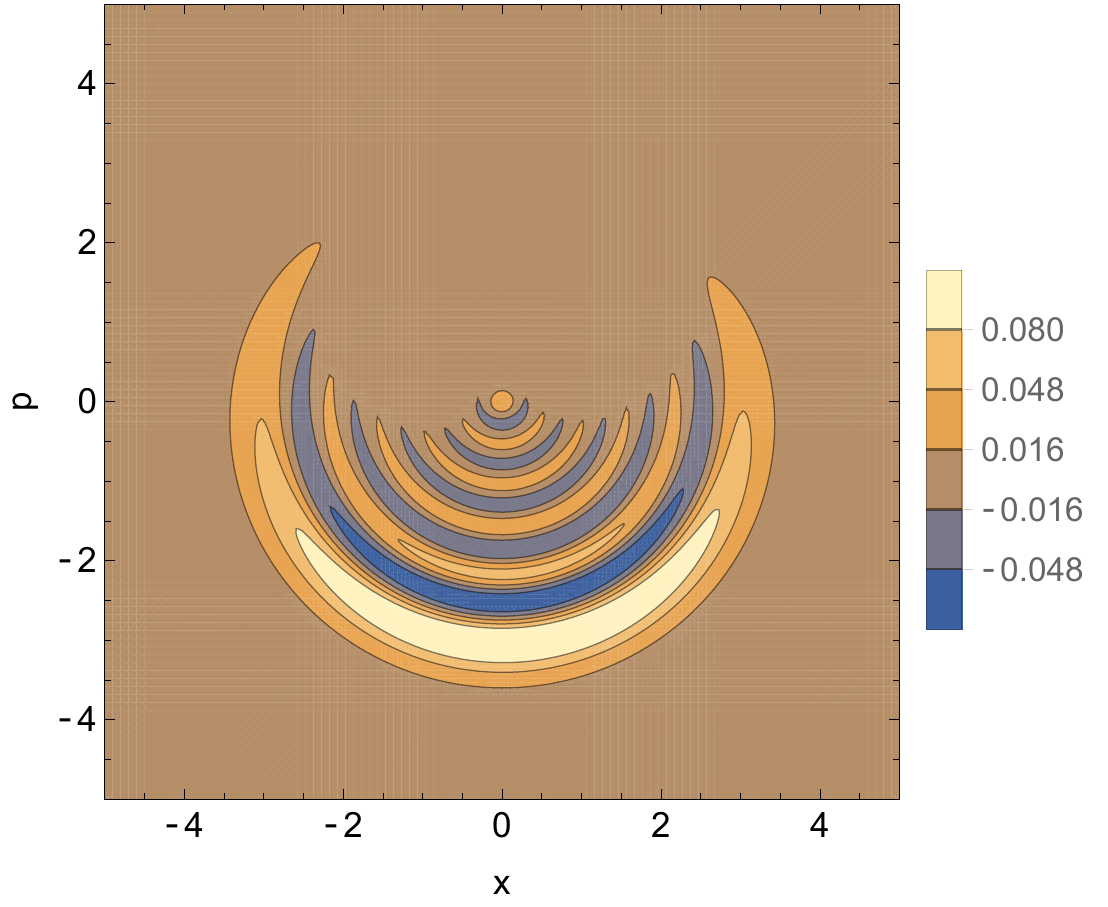}
\includegraphics[width=\columnwidth]{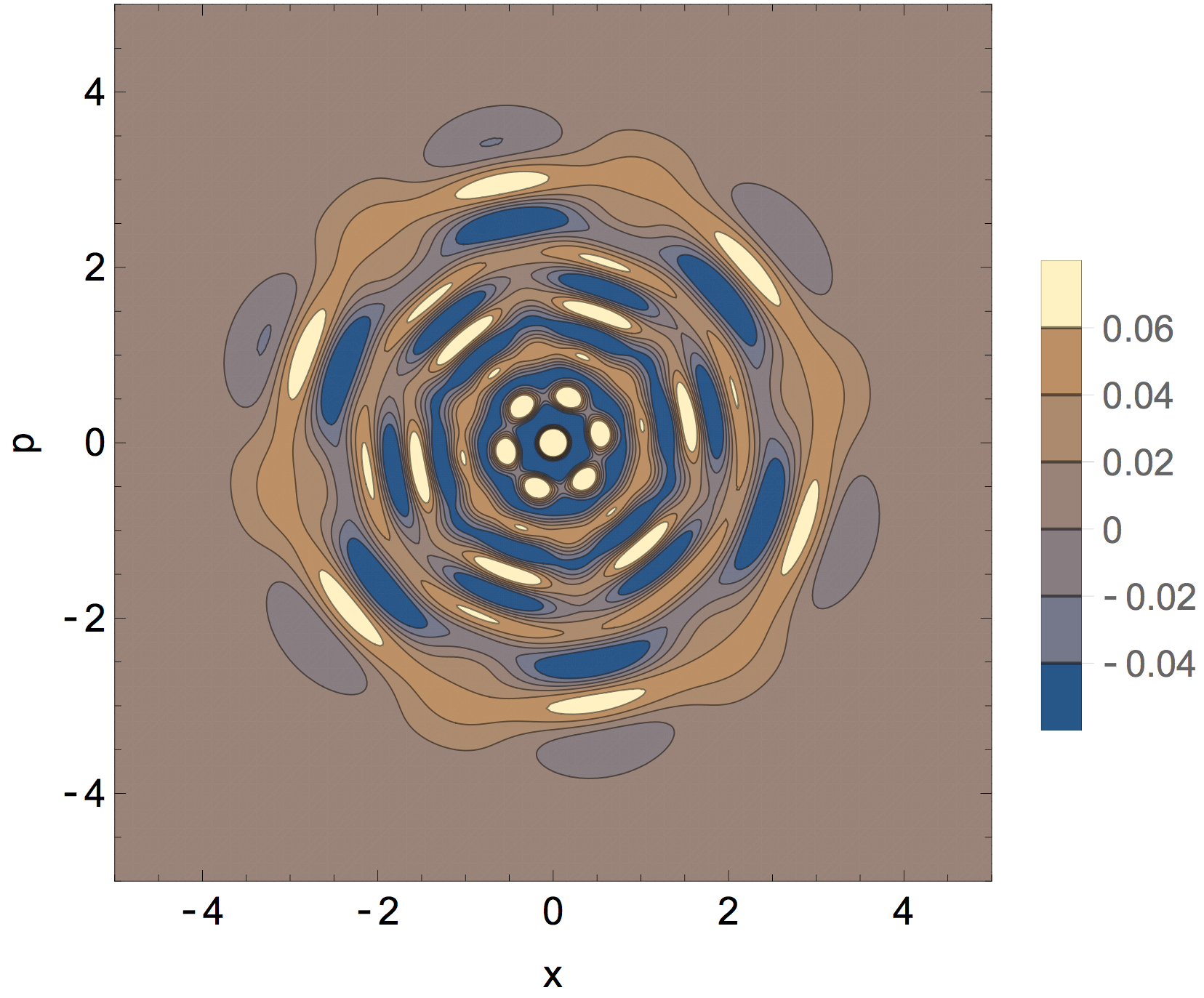}
\caption{(color online) As the number-phase entanglement increases, the Wigner function of the state smoothly changes from a coherent state, to an approximate quadrature-squeezed state (as in our proposal), to the Wigner function shown in \textbf{(a)}, to the Wigner function shown in \textbf{(b)}. The symmetry displayed in \textbf{(b)} occurs because the phase rotations are larger than 2$\pi$. For these Wigner functions only, experimental imperfections other than a loss of $\eta=0.7$ in the first mode are not considered.  Both the first and second mode start in coherent states with $\alpha=\beta=\sqrt{10}$ and the Kerr strength in \textbf{(a)} is $\phi_0=2/5$ while $\phi_0=1$ in \textbf{(b)}.}
\label{wigner}
\end{figure}

Furthermore, a host of beautiful nonclassical states can be created by post-selection once phase rotations exceed 2$\pi$, as shown by the resulting Wigner functions shown in Figure \ref{wigner}. The usefulness of arbitrary non-Gaussian states is an area of ongoing research \cite{resource, resource2, resource3}.

\begin{figure}[htbp]
    \centering
    \includegraphics[width=.8\columnwidth]{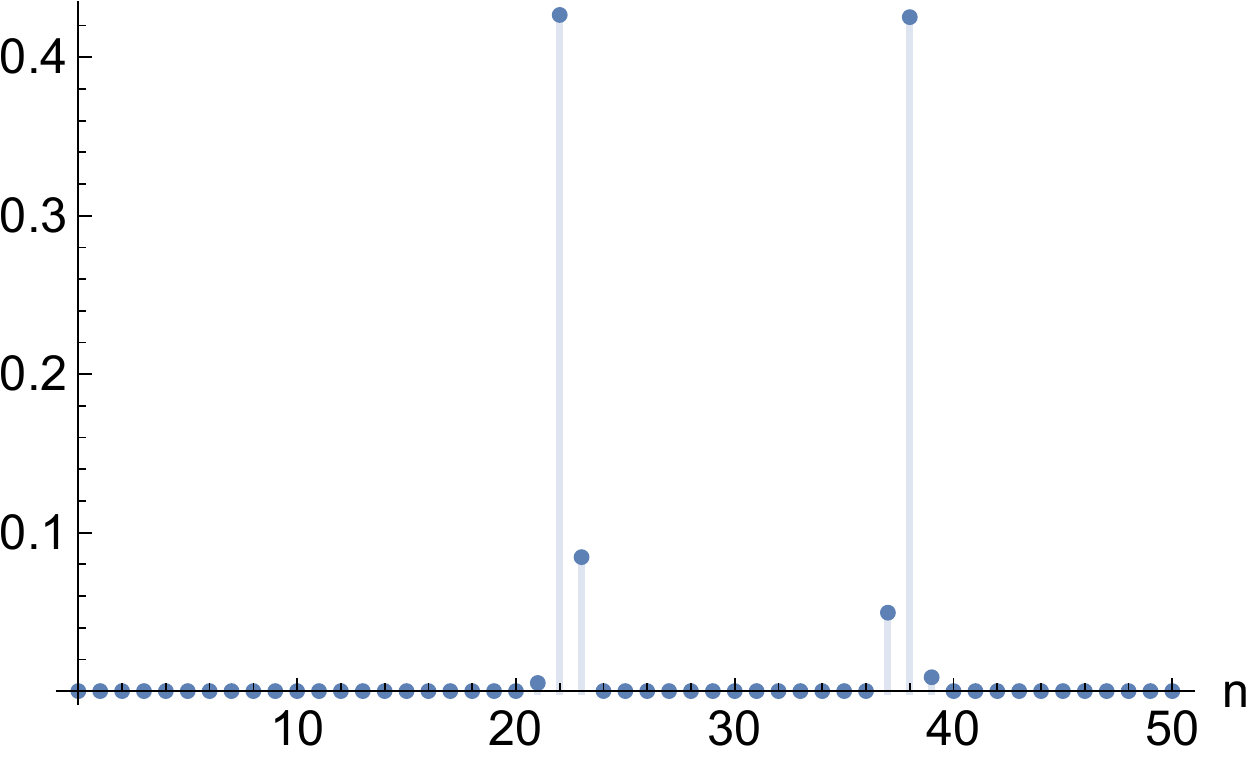}
    \caption{(color online) We set parameters $\alpha=\beta=\sqrt{30}$ and $\phi_0=2/5$ and exclude the influence of loss or noise.  By post-selecting on $\delta=-3.41+2.09i$, we obtain a state of the form $\sum^{\infty}_{n=0} c_n \ket{n}$, with amplitudes $c_n$ as shown in the plot.  Note that the average rotation arising from these parameters would be $\approx 5.71$ rad and our post-selection at $\approx 2.59$ rad corresponds to a shift of around $\pi$ as expected. As claimed above, the distribution is sharply peaked about two Fock states, differing in this case by 16 photons. By varying parameters, the peak sharpness and distance between peaks can be varied.}
    \label{twopeak}
\end{figure}

Another ambitious but interesting experiment would use large probe coherent states and large interaction strengths to get rotation angle differences on the order of 2$\pi$. Note that this does not require large single photon rotations if the uncertainty in the probe state photon number is large enough (roughly, $\Delta n \phi_0 \approx \pi$, where $\Delta n = \alpha$ is the photon number uncertainty for our input coherent state $\alpha$). Then, for proper parameter choices, the post-selection of a bin of phase rotations around $\pi$ is compatible with either the large-number \textit{or} small-number tails of $\alpha$'s photon number distribution, and the resultant state after post-selection is the coherent superposition of two peaks in photon number with quite different means. In the ideal case, they would differ by roughly $2 \Delta n $. For careful parameter choices, it is easy to see that we can create good approximations to the state $\frac{1}{\sqrt{2}}(\ket{m + \Delta n} + \ket{m - \Delta n})$ for some mean $m$, an example of which is seen in Figure \ref{twopeak}. For experiments in the microwave regime, the creation of these states might well be possible, since interactions strengths are roughly large enough \cite{exp3}. This state is of fundamental interest, can be used to create cat states, and can enhance experiments which rely on differences in photon number, e.g. optomechanical decoherence tests. 

Many experiments take direct advantage of the difference in photon numbers, typically using states of the form $\frac{1}{\sqrt{2}}(\ket{0} + \ket{1})$. In these cases, the state we have just proposed gives an immediate advantage. A simple example of this advantage arises in the optomechanical experiment described in \cite{towards}. Schematically, a superposition state $\frac{1}{\sqrt{2}}(\ket{0} + \ket{1})$ of two photon numbers is entangled via radiation pressure with the phonon coherent state occupation $\kappa(t)$ of a mesoscopic mechanical oscillator. The resulting micro-macro entangled state $\frac{1}{\sqrt{2}}(\ket{0}\ket{0} +\ket{1}\ket{\kappa(t)})$ is ideally suited for testing novel decoherence mechanisms; however, current experiments are limited by damping and achievable optomechanic interaction strengths $g$. One alternative to increasing $g$ is to start the protocol with our more exotic $\frac{1}{\sqrt{2}}(\ket{N} + \ket{M})$ state. To see how this would work, consider the usual optomechanical interaction Hamiltonian in the obvious interaction picture:
\begin{equation}
H_{int} = g(\hat{b} + \hat{b}^{\dagger})\hat{a}^{\dagger} \hat{a} \rightarrow g(\hat{b} e^{-i \omega_m t} + \hat{b}^{\dagger} e^{i \omega_m t})\hat{a}^{\dagger} \hat{a},
\end{equation}
where $\hat{a}$ and $\hat{b}$ label the photon and phonon modes respectively, and $\omega_m$ is the mechanical frequency. The time evolution (to first order in the Magnus expansion \cite{magnus}) of the interacting photon $\frac{1}{\sqrt{2}}(\ket{N}_a + \ket{M}_a)$ and phonon $\ket{0}_b$ states then becomes 
\begin{align*}
&e^{-i \int^t_0 H_I(t')dt'}\frac{1}{\sqrt{2}}(\ket{N}_a + \ket{M}_a)\ket{0}_b \\&= \frac{1}{\sqrt{2}}(\ket{N}_a e^{\frac{N g}{\omega_m}[(1-e^{-i\omega_m t})\hat{b} + h.a.]} \ket{0}_b \\& \; \; \; \; \; \; + \ket{M}_a e^{\frac{Mg}{\omega_m}[(1-e^{-i\omega_m t})+ h.a.]}  \ket{0}_b) \\&= \frac{1}{\sqrt{2}}(\ket{N}_a \ket{N \kappa(t)}_b + \ket{M}_a \ket{M \kappa(t)}_b)
\end{align*}
with phonon coherent state amplitude $\kappa(t) = \frac{4g}{\omega_m}\sin^2(\omega_m t/2)$. To create as entangled a state as possible, one desires that the coherent states in the two branches of the entangled state are as orthogonal as possible. The overlap
\begin{equation}
e^{-|M-N|^2 |\kappa(t)|^2}
\end{equation}  
has the same scaling with $g$ and $M-N$, so use of our exotic state is equivalent to increasing the interaction strength by a corresponding factor (compared with previous proposals, all of which use $M = 1$, $N = 0$). While creation of such a state is again very ambitious in the optical regime, the same can be said for optomechanics proposals in the strongly interacting regime, and it remains to be seen which toolbox will improve more rapidly. Using other experimental setups, such as Rydberg atoms or Josephson junctions, phase shifts of close to $\pi$ have been demonstrated \cite{Ryd2, exp3}. However, in the case of Rydberg atoms, the nonlinear interaction is saturated for a single photon \cite{Ryd}.

\section{Conclusions and Outlook}
A great deal of previous research has considered harnessing the entanglement created by interactions between two coherent states in a cross-Kerr medium for useful and interesting applications. We extend this body of work by analyzing a specific experiment using a weak cross-Kerr interaction to create novel bright, number-squeezed states. By simulating all realistic experimental conditions (some of which were notably absent from past models), we find that such an experiment is almost feasible with current technology, but requires improved isolation from background counts. We suggest an optimal set of parameters for a proof-of-principle experiment. Finally, we explore ambitious extensions of our scheme which would yield a variety of interesting, nonclassical resource states, but could only be realized with dramatic improvements in cross-Kerr interaction strengths. Some of our novel techniques would allow for creation of highly nonclassical states that have not been previously noted. We propose the use of these states in quantum optomechanics as a means of enhancing the generation of interesting phonon states.

\section{Acknowledgements}
The original idea for our experimental proposal was conceived in collaboration with Aephraim Steinberg, Josiah Sinclair, Matin Hallaji, and Greg Dmochowski; we thank them for useful discussions towards making a realistic model. Finally, we thank Nicol\'as Quesada for helpful input and fruitful discussions.

\section{Disclosure statement}
No potential conflict of interest was reported by the authors.

\section{Funding}
D.S., K.M., and D.F.V.J acknowledge support from the Natural Sciences and Engineering Research Council of Canada.

\bibliographystyle{tfp}
\bibliography{refs}

\begin{thebibliography}{10}
\providecommand{\url}[1]{\normalfont{#1}}
\providecommand{\urlprefix}{}

\bibitem{exp}
Fushman, I.; Englund, D.; Faraon, A.; Stoltz, N.; Petroff, P.; Vu{\v
  c}kovi{\'c}, J. Controlled Phase Shifts with a Single Quantum Dot,
  \emph{Science}  \textbf{2008}, \emph{320}~(5877), 769--772.
  \urlprefix\url{http://science.sciencemag.org/content/320/5877/769}.

\bibitem{exp2}
Venkataraman, V.; Saha, K.; Gaeta, A.L. Phase modulation at the few-photon
  level for weak-nonlinearity-based quantum computing, \emph{Nat Photon}
  \textbf{2013}, \emph{7}~(2), 138--141.
  \urlprefix\url{http://dx.doi.org/10.1038/nphoton.2012.283}.

\bibitem{exp3}
Kirchmair, G.; Vlastakis, B.; Leghtas, Z.; Nigg, S.E.; Paik, H.; Ginossar, E.;
  Mirrahimi, M.; Frunzio, L.; Girvin, S.M.; Schoelkopf, R.J. Observation of
  quantum state collapse and revival due to the single-photon Kerr effect,
  \emph{Nature}  \textbf{2013}, \emph{495}~(7440), 205--209.
  \urlprefix\url{http://dx.doi.org/10.1038/nature11902}.

\bibitem{amir}
Feizpour, A.; Hallaji, M.; Dmochowski, G.; Steinberg, A.M. Observation of the
  nonlinear phase shift due to single post-selected photons, \emph{Nat Phys}
  \textbf{2015}, \emph{11}~(11), 905--909. Letter,
  \urlprefix\url{http://dx.doi.org/10.1038/nphys3433}.

\bibitem{QC}
Lloyd, S.; Braunstein, S.L. Quantum Computation over Continuous Variables,
  \emph{Phys. Rev. Lett.}  \textbf{1999}, \emph{82}, 1784--1787.
  \urlprefix\url{http://link.aps.org/doi/10.1103/PhysRevLett.82.1784}.

\bibitem{kerrgate}
Lin, Q.; Li, J. Quantum control gates with weak cross-Kerr nonlinearity,
  \emph{Phys. Rev. A}  \textbf{2009}, \emph{79}, 022301.
  \urlprefix\url{http://link.aps.org/doi/10.1103/PhysRevA.79.022301}.

\bibitem{squeez}
Kitagawa, M.; Yamamoto, Y. Number-phase minimum-uncertainty state with reduced
  number uncertainty in a Kerr nonlinear interferometer, \emph{Phys. Rev. A}
  \textbf{1986}, \emph{34}, 3974--3988.
  \urlprefix\url{http://link.aps.org/doi/10.1103/PhysRevA.34.3974}.

\bibitem{highly}
Tyc, T.; Korolkova, N. Highly non-Gaussian states created via cross-Kerr
  nonlinearity, \emph{New Journal of Physics}  \textbf{2008}, \emph{10}~(2),
  023041. \urlprefix\url{http://stacks.iop.org/1367-2630/10/i=2/a=023041}.

\bibitem{kerrgen}
Zhang, Z.M.; Khosa, A.H.; Ikram, M.; Zubairy, M.S. Generating entangled states
  of continuous variables via cross-Kerr nonlinearity, \emph{Journal of Physics
  B: Atomic, Molecular and Optical Physics}  \textbf{2007}, \emph{40}~(10),
  1917. \urlprefix\url{http://stacks.iop.org/0953-4075/40/i=10/a=024}.

\bibitem{kerrgen2}
Yurke, B.; Stoler, D. The dynamic generation of Schr{\"o}dinger cats and their
  detection, \emph{Physica B+C}  \textbf{1988}, \emph{151}~(1), 298 -- 301.
  \urlprefix\url{http://www.sciencedirect.com/science/article/pii/0378436388901817}.

\bibitem{tian}
Wang, T.; Lau, H.W.; Kaviani, H.; Ghobadi, R.; Simon, C. Strong micro-macro
  entanglement from a weak cross-Kerr nonlinearity, \emph{Phys. Rev. A}
  \textbf{2015}, \emph{92}, 012316.
  \urlprefix\url{http://link.aps.org/doi/10.1103/PhysRevA.92.012316}.

\bibitem{macro}
Tara, K.; Agarwal, G.S.; Chaturvedi, S. Production of Schr\"odinger macroscopic
  quantum-superposition states in a Kerr medium, \emph{Phys. Rev. A}
  \textbf{1993}, \emph{47}, 5024--5029.
  \urlprefix\url{http://link.aps.org/doi/10.1103/PhysRevA.47.5024}.

\bibitem{weakkerrgen2}
Jeong, H. Using weak nonlinearity under decoherence for macroscopic
  entanglement generation and quantum computation, \emph{Phys. Rev. A}
  \textbf{2005}, \emph{72}, 034305.
  \urlprefix\url{http://link.aps.org/doi/10.1103/PhysRevA.72.034305}.

\bibitem{QIP1}
Dell’Anno, F.; Siena, S.D.; Illuminati, F. Multiphoton quantum optics and
  quantum state engineering, \emph{Physics Reports}  \textbf{2006},
  \emph{428}~(2–3), 53 -- 168.
  \urlprefix\url{http://www.sciencedirect.com/science/article/pii/S0370157306000329}.

\bibitem{weakkerrloss1}
Mogilevtsev, D.; Tyc, T.c.v.; Korolkova, N. Influence of modal loss on quantum
  state generation via cross-Kerr nonlinearity, \emph{Phys. Rev. A}
  \textbf{2009}, \emph{79}, 053832.
  \urlprefix\url{http://link.aps.org/doi/10.1103/PhysRevA.79.053832}.

\bibitem{QC2}
Munro, W.J.; Nemoto, K.; Spiller, T.P. Weak nonlinearities: a new route to
  optical quantum computation, \emph{New Journal of Physics}  \textbf{2005},
  \emph{7}~(1), 137.
  \urlprefix\url{http://stacks.iop.org/1367-2630/7/i=1/a=137}.

\bibitem{tele}
Seshadreesan, K.P.; Dowling, J.P.; Agarwal, G.S. Non-Gaussian entangled states
  and quantum teleportation of Schrödinger-cat states, \emph{Physica Scripta}
  \textbf{2015}, \emph{90}~(7), 074029.
  \urlprefix\url{http://stacks.iop.org/1402-4896/90/i=7/a=074029}.

\bibitem{metrol}
Whittaker, R.; Erven, C.; Neville, A.; Berry, M.; O'Brien, J.L.; Cable, H.;
  Matthews, J.C.F. Quantum-enhanced absorption spectroscopy, 2015.

\bibitem{ND}
Munro, W.J.; Nemoto, K.; Beausoleil, R.G.; Spiller, T.P. High-efficiency
  quantum-nondemolition single-photon-number-resolving detector, \emph{Phys.
  Rev. A}  \textbf{2005}, \emph{71}, 033819.
  \urlprefix\url{http://link.aps.org/doi/10.1103/PhysRevA.71.033819}.

\bibitem{cat}
Ralph, T.C.; Gilchrist, A.; Milburn, G.J.; Munro, W.J.; Glancy, S. Quantum
  computation with optical coherent states, \emph{Phys. Rev. A}  \textbf{2003},
  \emph{68}, 042319.
  \urlprefix\url{http://link.aps.org/doi/10.1103/PhysRevA.68.042319}.

\bibitem{resource2}
Ghose, S.; Sanders, B.C. Non-Gaussian ancilla states for continuous variable
  quantum computation via Gaussian maps, \emph{Journal of Modern Optics}
  \textbf{2007}, \emph{54}~(6), 855--869.
  \urlprefix\url{http://dx.doi.org/10.1080/09500340601101575}.

\bibitem{entang}
Silberhorn, C.; Lam, P.K.; Wei\ss{}, O.; K\"onig, F.; Korolkova, N.; Leuchs, G.
  Generation of Continuous Variable Einstein-Podolsky-Rosen Entanglement via
  the Kerr Nonlinearity in an Optical Fiber, \emph{Phys. Rev. Lett.}
  \textbf{2001}, \emph{86}, 4267--4270.
  \urlprefix\url{http://link.aps.org/doi/10.1103/PhysRevLett.86.4267}.

\bibitem{distill}
Fiur\'a\ifmmode~\check{s}\else \v{s}\fi{}ek, J.; Mi\ifmmode~\check{s}\else
  \v{s}\fi{}ta, L.; Filip, R. Entanglement concentration of continuous-variable
  quantum states, \emph{Phys. Rev. A}  \textbf{2003}, \emph{67}, 022304.
  \urlprefix\url{http://link.aps.org/doi/10.1103/PhysRevA.67.022304}.

\bibitem{mandel}
Short, R.; Mandel, L. Observation of Sub-Poissonian Photon Statistics,
  \emph{Phys. Rev. Lett.}  \textbf{1983}, \emph{51}, 384--387.
  \urlprefix\url{http://link.aps.org/doi/10.1103/PhysRevLett.51.384}.

\bibitem{teich89}
Teich, M.C.; Saleh, B.E.A. Squeezed state of light, \emph{Quantum Optics:
  Journal of the European Optical Society Part B}  \textbf{1989}, \emph{1}~(2),
  153. \urlprefix\url{http://stacks.iop.org/0954-8998/1/i=2/a=006}.

\bibitem{heter}
Walker, N. Quantum Theory of Multiport Optical Homodyning, \emph{Journal of
  Modern Optics}  \textbf{1987}, \emph{34}~(1), 15--60.
  \urlprefix\url{http://dx.doi.org/10.1080/09500348714550131}.

\bibitem{matin}
Sinclair, J.; Steinberg, A.M.; Hallaji, M.; Dmochowski, G. private
  communication, 2015.

\bibitem{Williamson36}
Williamson, J. On the Algebraic Problem Concerning the Normal Forms of Linear
  Dynamical Systems, \emph{American Journal of Mathematics}  \textbf{1936},
  \emph{58}~(1), 141--163. \urlprefix\url{http://www.jstor.org/stable/2371062}.

\bibitem{weedbrook}
Weedbrook, C.; Pirandola, S.; Garc\'{\i}a-Patr\'on, R.; Cerf, N.J.; Ralph,
  T.C.; Shapiro, J.H.; Lloyd, S. Gaussian quantum information, \emph{Rev. Mod.
  Phys.}  \textbf{2012}, \emph{84}, 621--669.
  \urlprefix\url{http://link.aps.org/doi/10.1103/RevModPhys.84.621}.

\bibitem{zou}
Zou, X.T.; Mandel, L. Photon-antibunching and sub-Poissonian photon statistics,
  \emph{Phys. Rev. A}  \textbf{1990}, \emph{41}, 475--476.
  \urlprefix\url{http://link.aps.org/doi/10.1103/PhysRevA.41.475}.

\bibitem{gerry}
Gerry, C.; Knight, P. \emph{Introductory Quantum Optics}; Cambridge University
  Press, 2005; \urlprefix\url{https://books.google.ca/books?id=CgByyoBJJwgC}.

\bibitem{comm}
Caves, C.M.; Drummond, P.D. Quantum limits on bosonic communication rates,
  \emph{Rev. Mod. Phys.}  \textbf{1994}, \emph{66}, 481--537.
  \urlprefix\url{http://link.aps.org/doi/10.1103/RevModPhys.66.481}.

\bibitem{bosonic}
Michael, M.H.; Silveri, M.; Brierley, R.T.; Albert, V.V.; Salmilehto, J.;
  Jiang, L.; Girvin, S.M. New Class of Quantum Error-Correcting Codes for a
  Bosonic Mode, \emph{Phys. Rev. X}  \textbf{2016}, \emph{6}, 031006.
  \urlprefix\url{http://link.aps.org/doi/10.1103/PhysRevX.6.031006}.

\bibitem{interf}
Holland, M.J.; Burnett, K. Interferometric detection of optical phase shifts at
  the Heisenberg limit, \emph{Phys. Rev. Lett.}  \textbf{1993}, \emph{71},
  1355--1358.
  \urlprefix\url{http://link.aps.org/doi/10.1103/PhysRevLett.71.1355}.

\bibitem{resource}
Gottesman, D.; Kitaev, A.; Preskill, J. Encoding a qubit in an oscillator,
  \emph{Phys. Rev. A}  \textbf{2001}, \emph{64}, 012310.
  \urlprefix\url{http://link.aps.org/doi/10.1103/PhysRevA.64.012310}.

\bibitem{resource3}
Genoni, M.G.; Paris, M.G.A. Quantifying non-Gaussianity for quantum
  information, \emph{Phys. Rev. A}  \textbf{2010}, \emph{82}, 052341.
  \urlprefix\url{http://link.aps.org/doi/10.1103/PhysRevA.82.052341}.

\bibitem{towards}
Marshall, W.; Simon, C.; Penrose, R.; Bouwmeester, D. Towards Quantum
  Superpositions of a Mirror, \emph{Phys. Rev. Lett.}  \textbf{2003},
  \emph{91}, 130401.
  \urlprefix\url{http://link.aps.org/doi/10.1103/PhysRevLett.91.130401}.

\bibitem{magnus}
Magnus, W. On the exponential solution of differential equations for a linear
  operator, \emph{Communications on Pure and Applied Mathematics}
  \textbf{1954}, \emph{7}~(4), 649?673.

\bibitem{Ryd2}
Tiarks, D.; Schmidt, S.; Rempe, G.; D{\"u}rr, S. Optical $\pi$ phase shift
  created with a single-photon pulse, \emph{Science Advances}  \textbf{2016},
  \emph{2}~(4).
  \urlprefix\url{http://advances.sciencemag.org/content/2/4/e1600036}.

\bibitem{Ryd}
Sevin\ifmmode~\mbox{\c{c}}\else \c{c}\fi{}li, S.; Henkel, N.; Ates, C.; Pohl,
  T. Nonlocal Nonlinear Optics in Cold Rydberg Gases, \emph{Phys. Rev. Lett.}
  \textbf{2011}, \emph{107}, 153001.
  \urlprefix\url{http://link.aps.org/doi/10.1103/PhysRevLett.107.153001}.

\end{thebibliography}

\end{document}